\documentclass[lengthcheck,twocolumn,superscriptaddress,showpacs,10pt,prb,floatfix,amssymb]{revtex4}

\usepackage{graphicx}

\begin{document}
\newlength{\plotwidth}          
\setlength{\plotwidth}{8.3cm}   

\title{Shot noise in self-assembled InAs quantum dots}
\author{A.~Nauen}
\email{nauen@nano.uni-hannover.de}
\author{I.~Hapke-Wurst}
\author{F.~Hohls}
\author{U.~Zeitler}
\author{R.~J.~Haug}
\affiliation{Institut f\"ur Festk\"orperphysik, Universit\"at Hannover,
Appelstr. 2, D-30167 Hannover, Germany}

\author{K.~Pierz}
\affiliation{Physikalisch-Technische Bundesanstalt, Bundesallee 100,
D-38116 Braunschweig, Germany}

\date{\today}

\begin{abstract}
We investigate the noise properties of a GaAs-AlAs-GaAs tunneling
structure with embedded self-assembled InAs quantum dots in the
single-electron tunneling regime. We analyze the dependence of the
relative noise amplitude of the shot noise on bias voltage. We observe
a non-monotonic behaviour of the Fano-factor $\alpha$ with an average
value of $\alpha \approx 0.8$ consistent with the asymmetry of the
tunneling barriers. Reproducible fluctuations observed in $\alpha$ can
be attributed to the successive participation of more and more InAs
quantum dots in the tunneling current.
\end{abstract}
 \maketitle

Performing noise measurements on microscopic semiconductor devices
allows an insight into details of the transport process not accessible
by conventional conductance experiments. In the most simple system for
such an experiment, an ideal tunneling barrier, a frequency-independent
power spectrum of the current noise up to frequencies corresponding to
the transit time of the carriers is observed. This case, known as full
shot noise, is due to a Poissonian statistics of the individual
tunneling events and a totally uncorrelated flow of charge carriers
\cite{ambrozy}.

If an additional source of negative correlation due to the
Pauli-exclusion principle is introduced the noise amplitude was shown
to be reduced~\cite{Li1990, Liu1995}. In these experiments, changing
the ratio of the transmissivity through the two barriers of a
double-barrier resonant-tunneling structure resulted into a suppression
of the relative shot-noise amplitude compared to full shot noise.
Theoretical models for purely coherent transport~\cite{Chen1991} and
for classical sequential tunneling~\cite{Davies1992} have been
developed. Both yield identical results for the noise suppression since
shot noise generally is not sensitive to dephasing
~\cite{BlanterReview}. This suppression of the shot noise was also
observed for resonant tunneling in zero-dimensional
systems~\cite{birk1995, andre1}. Under certain circumstances a positive
correlation between individual tunneling events can even increase the
noise power above the full Poissonian shot noise
value~\cite{iannaccone1998, kuznetsov1998}.


In this paper we present noise measurements on self-assembled InAs
quantum dot (QD) systems. These samples form an ideal system for noise
experiments since they provide zero-dimensional states of microscopic
dimensions. Furthermore, it is possible to select individual QDs for
transport by applying different bias voltages between the source and
drain contacts \cite{narihiro1997, eaves1996, suzuki1997, wurst2000}.
Consequently, we are able to measure the noise spectra of a resonant
tunneling current through single zero-dimensional states.

What we find in the experiments is a modulation of the amplitude of the
shot noise normalized to full shot noise as a function of the applied
bias voltage $V_{SD}$. We can link this to resonant single-electron
tunneling through individual InAs QDs in agreement with theoretical
expectations.



\begin{figure}[tb] 
  \begin{center}
      \resizebox{\plotwidth}{!}{\includegraphics{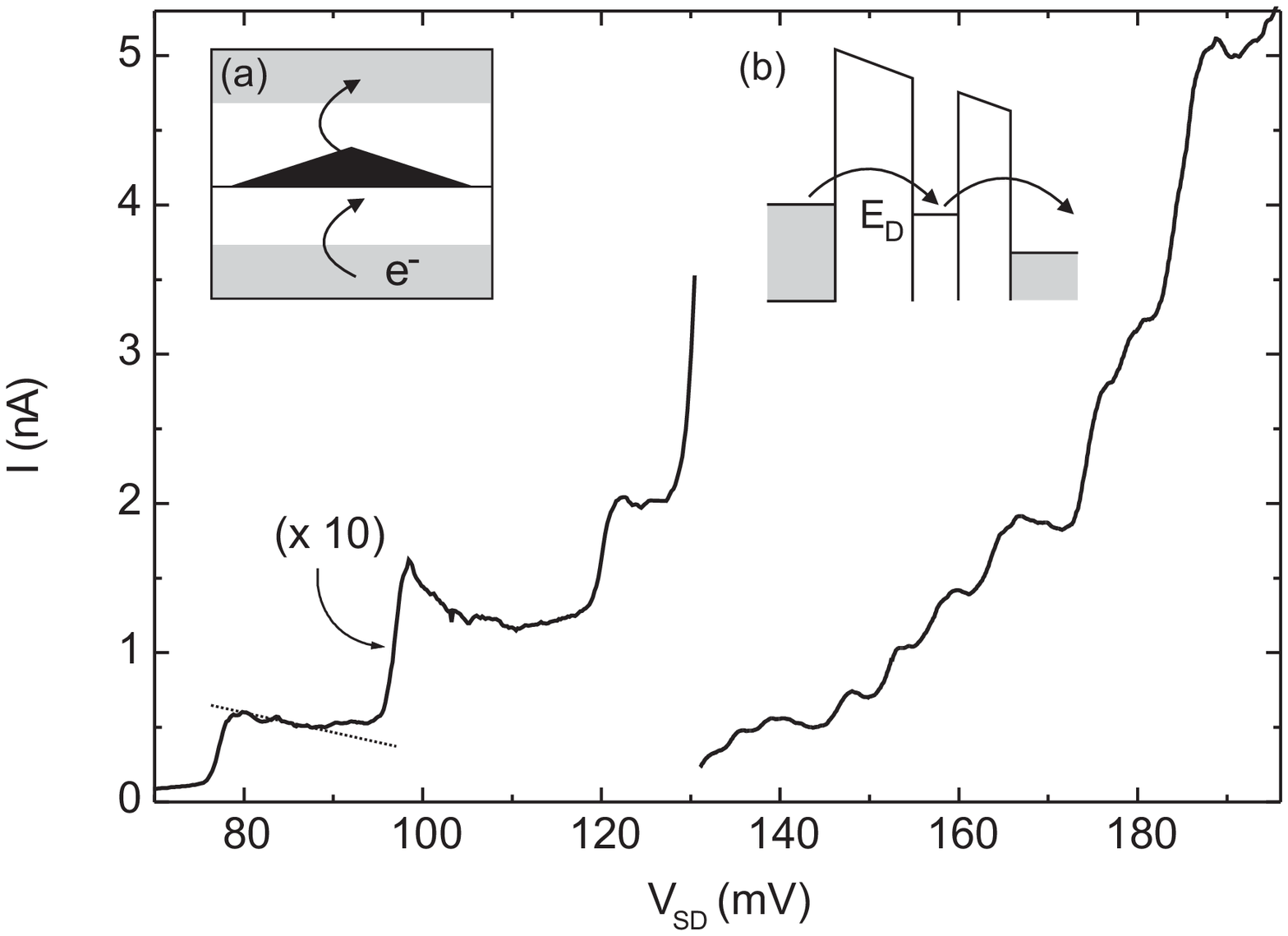}}
  \end{center}
  \caption{%
    Current-voltage characteristics of an InAs quantum dot array
    measured at $T=1.7$~K.
    The current for bias voltages below 130~mV is magnified by a factor of 10.
     The dotted line is a guide to the eye representing the expected
     current through a single dot.\\
     {\it Insets}: (a) Principle sample structure of an InAs QD (black)
     embedded in an AlAs barrier (white) between two GaAs electrodes
     (grey). The arrows mark the tunneling direction of the electrons.
     (b) Schematic profile of the band structure at positive bias where
         resonant-tunneling through a QD is observed.}
  \label{charakterisierung}
\end{figure}

The active part of our samples consists of a GaAs-AlAs-GaAs resonant
tunneling structure with embedded InAs QDs of 10-15~nm diameter and
3~nm height \cite{wurst1999}. These QDs are situated between two AlAs
barriers of nominally 4~nm and 6~nm thickness. The formation of
pyramidal QDs is due to the lattice mismatch between InAs and AlAs
resulting in a Stranski-Krastanov growth mode. About one million QDs
are placed randomly on the area of an etched diode structure of $40
\times 40\; \mu\mbox{m}^2$ area. A 15~nm undoped GaAs spacer layer and
a GaAs buffer with graded doping on both sides of the resonant
tunneling structure provide three-dimensional collector and emitter
electrodes. Connection to the active layer is realized by annealed
Au/Ge/Ni/Au contacts.

A schematic sample structure with one InAs QD embedded in an AlAs
barrier is sketched in the inset (a) of Fig.~\ref{charakterisierung}.
When applying a finite bias the zero-dimensional states of the QDs
inside of the AlAs barrier can be populated by electrons and a current
through the structure sets on, see Fig.~\ref{charakterisierung} (b). In
our experiments a positive bias voltage means tunneling first through
the base of a QD and out of the top. Due to the finite height of the
InAs QDs of approximately 3~nm their pinnacles penetrate into the top
AlAs barrier. This effectively reduces the width of the top barrier
below that of the nominally thinner barrier at the base of the QDs.
Therefore positive bias voltage corresponds to non-charging transport.

A typical current-voltage ($I$-$V$) characteristic is shown in
Fig.~\ref{charakterisierung}. We observe a step-like increase of the
current at bias voltages of 75, 95 and 120~mV. Each one of these
current steps corresponds to the emitter Fermi energy $E_F$ getting
into resonance with the ground states of different individual QDs. At
bias voltages lower than 70~mV no transport occurs since all the
zero-dimensional states inside the barrier lie above the emitter Fermi
energy $E_F$ and, consequently, electron flow is prohibited by the AlAs
barriers.


The negative slope of the first current plateaus is related to the
density of states of the three-dimensional emitter. Indeed, a
triangular-shaped $I$-$V$ curve as indicated by the dotted line is
expected for such a 3D-0D-3D tunneling diode. In particular with a
Fermi energy $E_F = 13.6$~meV and an energy-to-voltage conversion
factor of approximately 0.3 the current falls back to zero when the
distance to the onset voltage exceeds $\Delta V = 45$~mV
\cite{wurst2000}. Then the QD ground state with energy $E_D$ falls
below the conduction band edge $E_C$ of the emitter and no resonant
transport through this particular dot can take place anymore.

\begin{figure}[htb] 
  \begin{center}
      \resizebox{\plotwidth}{!}{\includegraphics{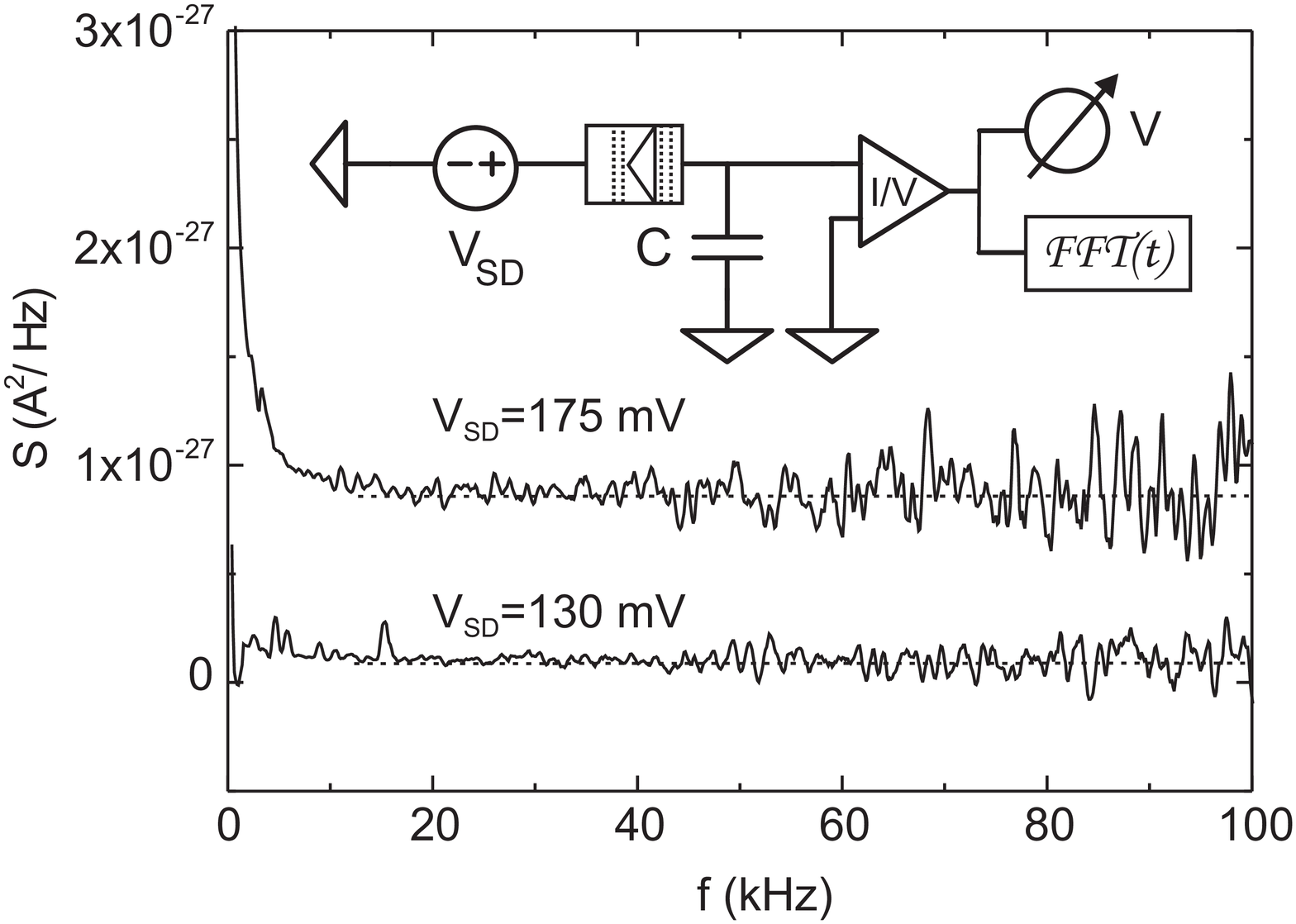}}
  \end{center}
  \caption{%
    Typical noise spectra of an array of self-assembled InAs quantum
    dots for two different bias voltages $V_{SD}$. The dashed horizontal lines
    are drawn to stress the absence of any frequency dependence above 20~kHz.
    \\
    {\it Inset}: Schematic of the experimental setup. A DC-voltage
    source drives a current through the device which
    is fed into a low-noise current amplifier by means of a
    low-capacitance cable with $C \approx 10$~pF. The noise spectra are
    acquired by a fast Fourier-transform analyzer (FFT). Additionally, the
    DC-part of the current is monitored.
    }
  \label{aufbau}
\end{figure}

The noise experiments are performed in a $^{4}$He bath cryostat with a
variable temperature insert. The sample is always immersed in liquid
helium. This allows for temperatures between 1.4~K and 4~K under stable
conditions.

The bias voltage $V_{SD}$ is applied between the source and drain
electrodes by means of a filtered DC-voltage source. The noise signal
is detected by a low-noise current amplifier in a frequency range from
0 to 100~kHz. The use of a current amplifier requires small capacitive
loading by the external circuit. Accordingly, electrical connection to
the sample is provided by a low-capacitance line  with $C \approx
10$~pF (see Fig.~\ref{aufbau}, inset).

The amplifier output is fed into a fast Fourier-transform analyzer
(FFT) to extract the noise spectra and in parallel into a voltmeter for
measuring the total DC-current through the sample. The amplifier and
the input stages of the FFT have been tested for linearity in the range
of interest, overall calibration has been verified by measuring the
thermal noise of thick film resistors.

By analyzing the frequency spectra we find that shot noise is present
above a certain frequency, as shown in Fig.~\ref{aufbau}. The spectra
have been smoothed using a boxcar average. The fluctuations of the
signal increase with frequency because of the capacitive loading of the
current amplifier. The superior signal-to-noise ratio at a bias voltage
$V_{SD}=130$~mV is due to a higher integration time. At frequencies
below $\sim 20$~kHz we find an additional $1/f$-noise contribution.

\begin{figure}[t] 
  \begin{center}
      \includegraphics[width=0.8\plotwidth]{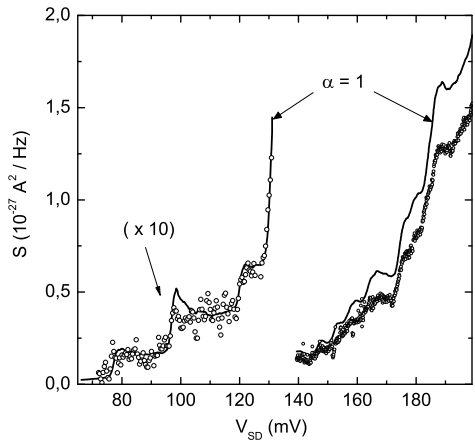}
  \end{center}
  \caption{%
    Measured shot noise (open circles) of the InAs QD diode. For comparison the calculated
    full shot noise ($\alpha = 1$) is plotted (black curve).
    The data for bias voltages smaller than 130~mV have been scaled
    with a factor of 10. Additionally,
    a 7-point boxcar average has been applied (the voltage
    resolution of the original data is $\Delta V_{SD} = 0.2$~mV).
    }
  \label{noise}
\end{figure}

For characterizing the relative amplitude of the shot noise we use the
dimensionless Fano factor $\alpha$, being defined as the ratio $\alpha
= S / S_{poisson}$ between full Poissonian shot noise $S_{poisson}=2eI$
(for $eV_{SD} \gg k_B T$) and the measured noise power density $S$,
with the total current $I$ and the electron charge $e$. Consequently,
full shot noise corresponds to $\alpha =1$. For the purpose of
extracting $\alpha$ from the spectra we average above the cut-off
frequency of the 1/f noise, thus increasing the signal to noise ratio.
The measured shot noise $S$ as a function of bias voltage $V_{SD}$ is
shown in Fig.~\ref{noise}. For comparison we plot the calculated full
shot noise $\alpha = 1$.


We consider the bias range where the first three QD states get into
resonance with the emitter Fermi energy $E_F$ as depicted in
Fig.~\ref{noise}. By averaging current and noise on the first plateau
we find a Fano factor $\alpha \approx 0.8$. The suppression of shot
noise for on-resonance tunneling is well understood theoretically (see
\cite{BlanterReview} and references therein): The value of the Fano
factor $\alpha$ is directly linked to the transmissivity aspect of left
and right barrier:
\begin{equation}
    \alpha =
    \frac{\Gamma_L^2+\Gamma_R^2}{\left(\Gamma_L+\Gamma_R\right)^2}\;.
    \label{formel}
\end{equation}
$\Gamma_L$ and $\Gamma_R$ are the partial decay widths of the resonant
state $E_D$. For symmetric barriers ($\Gamma_L = \Gamma_R$) the
suppression would become $\alpha = 1/2$. Responsible for this is the
Pauli exclusion principle: For symmetric barriers the tunneling of an
electron from the emitter into the QD is forbidden as long as the QD
state $E_D$ is occupied. This anti-correlates successive tunneling
events.  However, in the asymmetric case (e.g. $\Gamma_L \gg \Gamma_R$)
full poissonian shot noise ($\alpha = 1$) would be recovered, since
then the transport will be controlled by one barrier solely.

To get a more quantitative insight we use the textbook formula for the
transmission coefficient $T$ of a rectangular tunneling barrier
neglecting any influence of the applied bias voltage on the shape of
the barrier potential:
\begin{equation}
    T(E)=\frac{1}{1+
            \left\{1+
                \left(
                    \frac{1}{4}
                        \left(
                            \frac{\kappa}{k}-\frac{k}{\kappa}
                        \right)^2
                \right)
            \right\}
            \sinh^2(\kappa a)}
        \; ,
    \label{Tformel}
\end{equation}
with $\kappa=\sqrt{2m^* (V_0-E)}/\hbar$ and $k=\sqrt{2m^* E}/\hbar$,
the barrier width $a$ and height $V_0 = 1.05\; \mbox{eV}-E_F$, the
effective electron mass $m^*$ and the energy $E$ of the tunneling
emitter electrons.  This is justified since the AlAs barriers in our
sample are narrow but high. The transmission coefficients $T_{L,R}$ of
the left and right barrier are linked to the partial decay widths
$\Gamma_{L,R}$ via $T_{L,R} = \hbar \nu\; T_{L,R}$ with the attempt
rate $\nu = \sqrt{2 E_{D}/m^*}/w$ of the resonant QD state $E_{D}$ and
$w$ the distance between both barriers. Since $\nu$ is identical for
both barriers Eq.~(\ref{formel}) transforms into
\begin{equation}
    \alpha =
    \frac{T_L^2+T_R^2}{\left(T_L+T_R\right)^2}\;.
    \label{fanoT}
\end{equation}
We take the thickness of the tunneling barrier at the base of the QD
$a_{L}=4$~nm as fixed. From the measured Fano factor $\alpha \approx
0.8$ we extract from Eq.~(\ref{fanoT}) the ratio $T_{R}/T_{L} = 8.4$ at
a bias voltage of 0.1~V. We can now use Eq.~(\ref{Tformel}) to
calculate the thickness of the second barrier as $a_{R} = 3.2$~nm. The
top AlAs barrier has been grown with a nominal thickness of 6~nm and
the average height of an InAs QD is 3~nm. Hence, the calculated value
of $a_R = 3.2$~nm is in good agreement with the expected barrier
thickness of 3~nm.

\begin{figure}[t] 
  \begin{center}
      \includegraphics[width=\plotwidth]{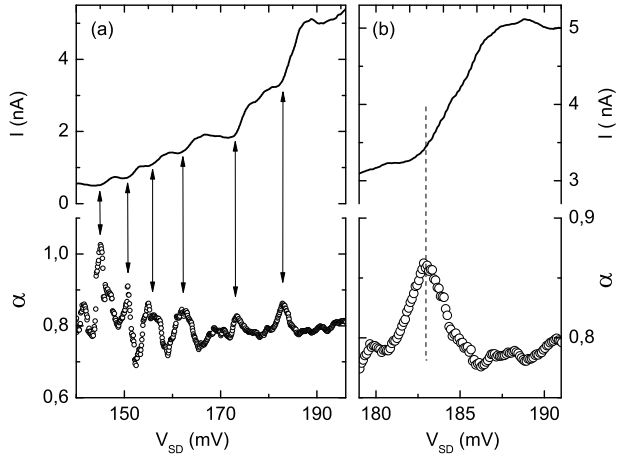}
  \end{center}
  \caption{%
    (a) $I$-$V$ characteristics of the sample at higher bias
    voltage (upper plot) in comparison with the
    measured Fano factor $\alpha$ (lower plot) for non-charging direction.
    The temperature was 1.7~K.
    The shot noise increases at every onset of a current plateau
    as indicated by the arrows. The plotted set of data for $\alpha$ has been smoothed using a 15-point boxcar
    average (the voltage resolution of the original data is $\Delta V_{SD} = 0.1$~mV).\\
    (b) Detailed plot of current (solid line) and corresponding Fano factor
    (open circles).
    }
  \label{fano}
\end{figure}

Up to now we have dealt with the bias regime where very few single QDs
determine the transport. If the bias voltage is increased above 130~mV
the energy separation between consecutive QD states is narrowed
considerably (Fig.~\ref{noise}) compared to the onset region.
Therefore, the number of QD states contributing to the current reaches
the order of ten at 190~mV. In this multiple-QD tunneling regime we
observe a non-monotonic behaviour of the Fano factor with the noise
being sub-Poissonian ($\alpha < 1$) on resonance. In between the
current plateaus we find increasing noise. However, there $\alpha$ is
found to be still sub-poissonian. For better insight into this
observation we have plotted the corresponding Fano factor $\alpha$ in
comparison with the $I$-$V$ characteristics in Fig.~\ref{fano}~(a).

First, we will concentrate on the increase of the Fano factor $\alpha$
at the onset of every current plateau (see Fig.~\ref{fano}(a)). A
detailed view of the Fano factor at a current step is plotted in
Fig.~\ref{fano}~(b). Full Poissonian shot noise is present in a stream
of electrons that is totally uncorrelated. The suppression of the shot
noise for on-resonance transport as predicted by Eq.~(\ref{formel}) is
due to a negative correlation in the electron flow induced by the Pauli
exclusion principle: As one electron enters the resonant QD state the
tunneling of further electrons into that state is forbidden. The
resulting dependence between consecutive tunneling events suppresses
the noise.

When a QD level starts coming into resonance only a few electrons from
the high energy tail of the Fermi distribution contribute to the
resonant current. This implies that the Pauli principle does not play a
role as long as an electron in the QD can tunnel out before the next
one is available in the emitter, in other words as long as the attempt
frequency in the emitter is smaller than the tunneling rate through the
second barrier. In such an uncorrelated case the Fano factor will
approach its Poissonian value $\alpha=1$, see Fig.~\ref{fano}~(b). If
the resonant state is moved further into the Fermi sea of the emitter,
the number of electrons available to resonant tunneling increases and
the Pauli principle sets in. The Fano factor starts becoming suppressed
until reaching its minimum on the current plateau.


Applying higher bias voltage we note in Fig.~\ref{fano}~(a) a reduction
of the overall modulation amplitude in the Fano factor $\alpha$,
eventually leading to an average suppression of $\alpha \approx 0.8$.
As we have stated above every single QD contributes to the total
current over a bias range of approximately $50$~mV. This implies that
e.g. the QD linked to the current step at $V_{SD} = 145$~mV in
Fig.~\ref{fano}~(a) transmits electrical current over most part of the
plotted bias range (up to $190$~mV) before its eigenstate moves below
the emitter conductance band edge $E_c$. The same holds for every
consecutive QD. Thus at a specific bias voltage $V_{SD}$ the total
current $I_{total}(V_{SD})$ is a sum of the currents $I_i$ through
every single QD : $I_{total}(V_{SD}) = \sum_{i=1}^{n(V_{SD})}
I_i(V_{SD})$ with $n(V_{SD})$ being the number of transmitting QDs.  As
indicated by the dotted line in Fig.~\ref{charakterisierung} each
current $I_i(V_{SD})$ features an approximately linear decreasing
behavior with increasing bias voltage.

From the theoretical models for on-resonance transport we know that the
suppression coefficient $\alpha_i$ is constant and its value is given
by Eq.~(\ref{formel}). We may speculate that in the multi-dot regime
the $\alpha_i$ are modified by additional mechanisms like interaction
effects. The fluctuations in the Fano factor minima may be traced to
that. Nevertheless, this implies that with increasing bias voltage we
have a growing number of QDs transmitting electrons with approximately
constant value for the Fano factor $\alpha$. Since we normalize the
measured noise on the {\it total} current this results in a reduction
of the effect of additional QDs coming into resonance.

In the limit of a large number of QDs being in resonance with the
emitter Fermi sea we expect any modulation of the Fano factor to
vanish, and its value to saturate at the value of the ensemble averaged
suppression: $\left<\alpha\right> = 1/n(V_{SD})\;
\sum_{i=1}^{n(V_{SD})} \alpha_i$. In fact, this is what we find for
bias voltages above 190~mV, where $n(190~\mbox{mV})$ is on the order of
ten: The measured value $\alpha \approx 0.8$ is once again in
reasonable agreement with the asymmetry of our tunneling structure (see
above).

Apart from the main features found in the noise and $I$-$V$
characteristics we observe a fine structure as shown exemplarily in
Fig.~\ref{fano}~(b) around 181~mV. Most likely this can be linked to
transport through QDs only weakly coupled to the emitter. Additionally,
these fine structures may be influenced by the fluctuations of the
local density of states in the emitter \cite{schmidt2001}.


To conclude, we have measured the shot noise of an array of
self-assembled InAs quantum dots (QDs). We observe a suppression of the
shot noise on current plateaus that is related to resonant tunneling
through the zero-dimensional ground state of a QD as predicted by
theory. In the step region between two current plateaus the noise
increases. This is because of the negligible influence of the Pauli
principle as long as the number of emitter electrons with the energy of
a specific QD state is sufficiently low. As more QD states lie between
the Fermi energy and the conduction band edge of the emitter we find a
suppression of the modulation amplitude of the shot noise.

We acknowledge financial support from DFG, BMBF, DIP and TMR.



\end{document}